\documentclass[a4paper]{jpconf}
\usepackage{graphicx}
\newcommand{\beq}{\begin{equation}}
\newcommand{\eeq}{\end{equation}}
\newcommand{\ba}{\begin{array}}
\newcommand{\ea}{\end{array}}
\newcommand{\gsim}   {\mathrel{\mathop{\kern 0pt \rlap
  {\raise.2ex\hbox{$>$}}}
\lower.9ex\hbox{\kern-.190em $\sim$}}}

\begin{document}

\title{Astroparticle Physics: Puzzles and Discoveries}

\author{V. Berezinsky}

\address{INFN, Laboratori Nazionali del Gran Sasso,
Assergi (AQ) 67010 Italy\\
and Institute for Nuclear Research, Moscow\\
}


\begin{abstract} %
Puzzles often give birth to the great discoveries, 
the false discoveries sometimes stimulate the exiting  
ideas in theoretical physics. The  historical examples of 
both are described in Introduction and in section ``Cosmological   
Puzzles''. From existing puzzles most attention is given to 
Ultra High Energy Cosmic Ray (UHECR) puzzle and to cosmological
constant problem. The 40-years old UHECR problem consisted in absence 
of the sharp steepening in spectrum of extragalactic cosmic rays  
caused by interaction with CMB radiation. This steepening is known 
as Greisen-Zatsepin-Kuzmin (GZK) cutoff. It is demonstrated here 
that the features of interaction of cosmic ray protons with CMB 
are seen now in the spectrum in the form of the dip and beginning
of the GZK cutoff. The most serious cosmological problem is caused 
by large vacuum energy of the known elementary-particle fields which 
exceeds at least by 45 orders of magnitude the cosmological vacuum  
energy. The various ideas put forward to solve this problem 
during last 40 years, have weaknesses and cannot be accepted as the 
final solution of this puzzle. The anthropic approach is discussed. 
\end{abstract}


\section{Introduction}
\begin{center}
        ALL GREAT DISCOVERIES IN ASTROPHYSICS\\
        APPEARED UNPREDICTABLY AS PUZZLES.\\
        WHAT WAS PREDICTED WAS NOT FOUND.\\
\end{center}
\vspace{5mm} Not many good things fall down on us from the sky, but
discoveries do. They arrive as the puzzles, and usually disappear as
the errors. But sometimes they become real, leaving behind the great
discoveries. I will give below a short list of astrophysical
discoveries of the last four decades, separating intuitively
astrophysics from cosmology.\\*[3mm]
{\it Quasars} were discovered in early 1960s as compact 
radio sources. G. Mathews and A. Sandage in 1960
identified radio source 3C48 with a stellar-like object. M. Schmidt
in 1963 deciphered the optical spectrum of quasar 3C273 assuming its
redshift, $z=0.158$. Surmounting resistance of sceptics, this
explanation moved the source to the distance of 630~Mpc and made its
luminosity uncomfortably large, $L \sim 10^{46}$~erg/s. It was a
puzzle, and many respectable astrophysicists spent years trying to
squeeze the source back into Galaxy. All of them failed, and in the
end the puzzling energy release resulted in the discovery of a {\it
black hole}, an object of general relativity.  \\ %
*[3mm] {\it Pulsars} were discovered first in 1967 by a student of
A. Hewish, Jocelyn Bell. She observed a puzzling periodicity of
radio-pulses from an unknown source. After short but intense
discussion of different possible sources, including extraterrestrial
civilizations and ``little green men'', the magnetized rotating
neutron stars, the pulsars, were found to be responsible. It opened
a new field of cosmic physics: {\em relativistic electrodynamics}.\\ %
*[3mm] {\it The atmospheric neutrino anomaly and the solar neutrino
problem} went along most difficult road  to the status
of discovery. The puzzling phenomenon in both cases was a neutrino
deficit as compared with calculations. The solar neutrino
experiments have been started by Ray Davis in 1960 in Brookhaven.
With time the solar-neutrino deficit raised, but scepticism of the
community raised too. Pushed first by Ray Davis and John Bahcall,
the solar neutrino problem moved like  a slow coach along a road of
three decades long. Fortunately, physics differs from democracy:
opinion of majority means usually less than that of ONE. In the end
the two obscure puzzles (solar and atmospheric neutrino deficits)
have turned  into discovery of the most fascinating phenomenon, {\it
neutrino oscillations}, theoretically predicted by Bruno Pontecorvo.
\\*[3mm]
{\it Supernova SN 1987a} became an elementary-particle
laboratory in the sky for a study of properties of neutrinos,
axions, majorons {\it etc}. Detection of neutrinos 
became a triumph of the theory: the number of detected neutrinos,
duration of the neutrino pulse and estimated neutrino luminosity
have appeared in agreement with theoretical prediction.
Gravitational collapse as a phenomenon providing the SN explosion
has been established ... and a puzzle appeared. It was
the  triumph of the incomplete theory, without rotation of
collapsing star. The asymmetric ring around SN 1987a implies that
the presupernova was a rotating star (it would be a surprise if
not!). Rotation should diminish the neutrino luminosity
or even change the collapsing scenario. Why then there is a
beautiful agreement between theory and observation? This problem
still expects its solution.
\subsection{Greatness of false discoveries}
\begin{center}
WHEN FIRST APPEARED THE PUZZLES LOOK WEAK.\\
SAVE YOUR TIME SAYING: IT'S RUBBISH.\\
IN 90$\%$ OF CASES YOU WILL BE RIGHT,\\
BUT YOU MAY MISS THE GREAT PHYSICS.
\end{center}
\vspace{3mm} False discoveries often have great impact on physics,
and {\it Cyg X-3 } is a famous example.\\*[1mm] Cyg
X-3 is a galactic binary system well studied in all types of
radiations, most notably in X-rays. In 80s many EAS arrays detected
from it the 4.8 hour periodic ``gamma-ray'' signal in VHE
(Very High Energy, $E\geq 1$~TeV) and UHE (Ultra High Energy, $E\geq
0.1-1$~PeV) ranges. The list of these arrays included Kiel, Haverah
Park, Fly's Eye, Akeno, Carpet-Baksan, Tien-Shan, 
Plateau Rosa, Durham, Ooty, Ohya, Gulmarg, Crimea,
Dugway, Whipple and others. Probably it is easy to say that there
was no single EAS array which claimed no-signal observation.
Additionally, some underground detectors (NUSEX, Soudan, MUTRON)
marginally observed high energy muon signal from the direction of
this source. Apart from the Kiel array, which claimed $6\sigma$
signal, the confidence level of detection was not high: $3-4
\sigma$.

In 1990 - 1991 two new-generation
detectors, CASA-MIA and CYGNUS, put the stringent upper limit to the signal
from Cyg X-3, which excluded early observations.

Apart from two lessons:
\begin{enumerate}
  \item good detectors are better than bad ones,
  \item ``$3\sigma$'' discoveries should not be trusted, even if many
detectors confirm them,
\end{enumerate}
experience of Cyg X-3 has taught us how to evaluate statistical
significance searching for periodic signals.

The false discovery of high energy radiation from Cyg X-3 had great
impact on theoretical high energy astrophysics, stimulating study of
acceleration in binary systems, production of high energy gamma and neutrino
radiation  and creation of high energy astrophysics with new
particles, such as light neutralinos, gluinos {\it etc}.
\section{Ultra High Energy Cosmic Ray (UHECR) Puzzle}
This is the oldest puzzle in physics which exists for 40 years, and
most probably now is close to being resolved. It appeared in 1966
together with prediction of the Greisen-Zatsepin-Kuzmin (GZK) cutoff
\cite{GZK}. As physical phenomenon the GZK cutoff is explained by
production of pions by UHE extragalactic protons
interacting with CMB photons. The energy of CMB photon in the
Lorentz system, where proton is at rest is $\Gamma$ times larger
than in the laboratory system (where $\Gamma$ is Lorentz factor of a
proton) and when this energy exceeds the mass of pion, the proton
energy loss becomes large, and the spectrum of UHE protons obtains
steepening which is (incorrectly but impressively!) called the ``GZK
cutoff''. For the diffuse spectrum the GZK feature (steepening)
starts at energy $(3 - 5)\times 10^{19}$~eV.
\begin{figure}[ht]
\centering
\includegraphics[width=8.0cm]{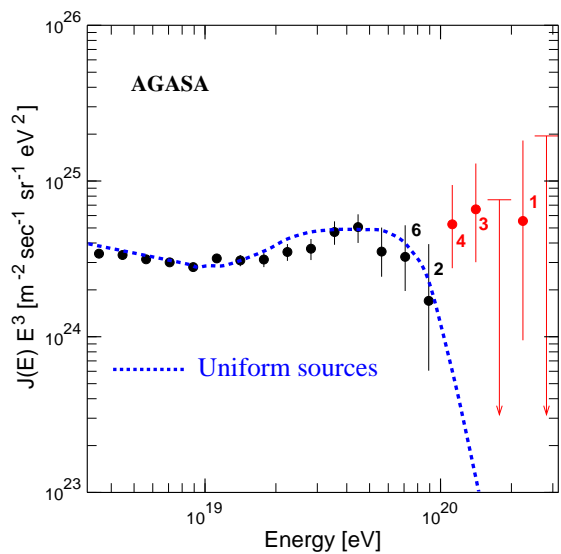}
\vspace{2mm} \caption{UHECR spectrum measured by AGASA detector in
1990 -1996 in comparison with calculated diffuse spectrum. The
excess of events over predicted GZK cutoff is 
clearly seen. At present ``AGASA excess'' includes 11 events
 at energy $E \geq 1\times 10^{20}$~eV. }
\label{fig:agasa96}
\end{figure}
The UHECR puzzle was born simultaneously with the prediction of the
GZK feature: already in 1966 there were known at least three events
with energies above the GZK cutoff. With time the number of these
particles increased and in 1994 the famous golden Fly's Eye event
\cite{FE} with energy $\sim 3\times 10^{20}$~eV was detected. The
energy of this event was determined very reliably and with the high
precision. It questioned the presence of the GZK cutoff. However,
most serious arguments against presence of the  GZK
feature were given by AGASA data \cite{AGASA}. In
Fig.~\ref{fig:agasa96} the excess of AGASA events over predicted GZK
cutoff is shown for data of 1996 - 1998 with 8 detected particles at
energy $E  \gsim 1\times 10^{20}$~eV, in 2004 this number reaches
reached 11 (for a review of observational data see
\cite{NaWa}). The GZK cutoff is predicted for extragalactic protons, but
already in 70s it was known that other conservative UHE signal
carriers, UHE nuclei, suffer the steepening of the spectrum
approximately at the same energy as protons \cite{BGZ,PSB}. The
other reasonable carriers, UHE photons, have too small absorption
length interacting with radio-photons (see \cite{radio}).\\
\subsection{New-physics solutions}
No viable astrophysical solution to the ``AGASA excess'' was found,
and the solutions with ``new physics'' have been proposed. I will
give here the short list of the new solutions without references
which can be found in \cite{Be2000}
\begin{itemize}
\item {\it Superheavy Dark Matter}. Long-lived Superheavy Dark Matter
Particles are accumulated in galactic halos. These particles are
naturally produced gravitationally at post-inflationary epoch and at
mass $m_x \sim 10^{13}$~GeV they provide CDM density as
observed by WMAP. These particles can be long-lived, with lifetime
exceeding the age of the Universe. Decays of these particles produce
UHECR without GZK cutoff (most of UHE particles come from Galactic
halo).
\item {\it Topological Defects} (TD). There are various mechanisms of
production of UHE particles by TD. In some cases TD become unstable
and decompose to constituent fields (superheavy Higgs and gauge
bosons), which then decay to ordinary particles. This mechanism
works for cusps and superconducting cosmic strings. In case of
monopoles and antimonopoles connected by strings, high energy
particles are produced at annihilation of monopole-antimonopole
pairs. The most promising candidates are necklaces and
monopole-antimonopole pairs connected by string. UHECR from TD 
have a spectrum with a soft GZK cutoff which does not
contradict observations.
\item {\it Resonant neutrinos}. Very high energy neutrinos are resonantly
absorbed by target neutrinos comprising Hot Dark Matter (HDM):
$\nu+{\bar \nu}_{HDM} \to Z^0 \to hadrons$. Very large flux of
primary neutrinos with superhigh energies is needed
for this hypothesis.
\item {\it Light gluino}. Light gluinos can be effectively produced
by TD or in pp-collisions in astrophysical sources. They weakly
degrade in energy interacting with microwave radiation. The
interaction of UHE light gluino with nucleons is similar to that of
UHE proton. Light gluino is disfavored by accelerator experiments.
\item {\it Strongly interacting neutrino}. In extra-dimension
theories, for example,
neutrino can have large cross-section of scattering off the nucleon.
In this case neutrino can be a carrier of UHE signal from remote
astrophysical sources.
\item {\it Lorentz invariance  breaking}. In this case for protons with
energies $10^{20}$~eV and higher, the c.m.\ energy could be not
enough for production of pions in collisions with microwave photons.
\end{itemize}
It is interesting that these new ideas, in particular, Superheavy
Dark Matter and Lorentz-invariance violation, first proposed in 1972
\cite{Kirzh} to solve UHECR puzzle, are developed now independently
of UHECR.
\subsection{Towards astrophysical solution}
GZK cutoff is nothing but a signature of UHE proton interaction with
CMB.

Are there some other signatures of the same interaction, and what they
tell us about UHECR problem?

Such signature is known since long time \cite{BG}: this is the {\em
dip\/}; the tiny feature in the proton spectrum left behind by
production of electron-positron pairs in collisions with CMB photons
($p+\gamma_{\rm CMB} \to p+e^++e^-$). This feature is seen better
when analyzed in terms of the modification factor \beq \eta (E)
=J_p(E)/J_p^{\rm unm}(E), \label{eq:eta} \eeq where $J_p(E)$ is
proton spectrum calculated with all energy losses included and
$J_p^{\rm unm}(E)$ is one calculated with adiabatic energy losses
(red-shift) only. The advantage of modification factor is its
model-independence \cite{BGG-pl,BGG-prd,Aloisio}, in contrast to GZK
cutoff which at $E \geq 1\times 10^{20}$~eV is strongly model
dependent.

\begin{figure}[t]
\begin{center}
   \begin{minipage}[ht]{54mm}
     \centering
     \includegraphics[width=53mm]{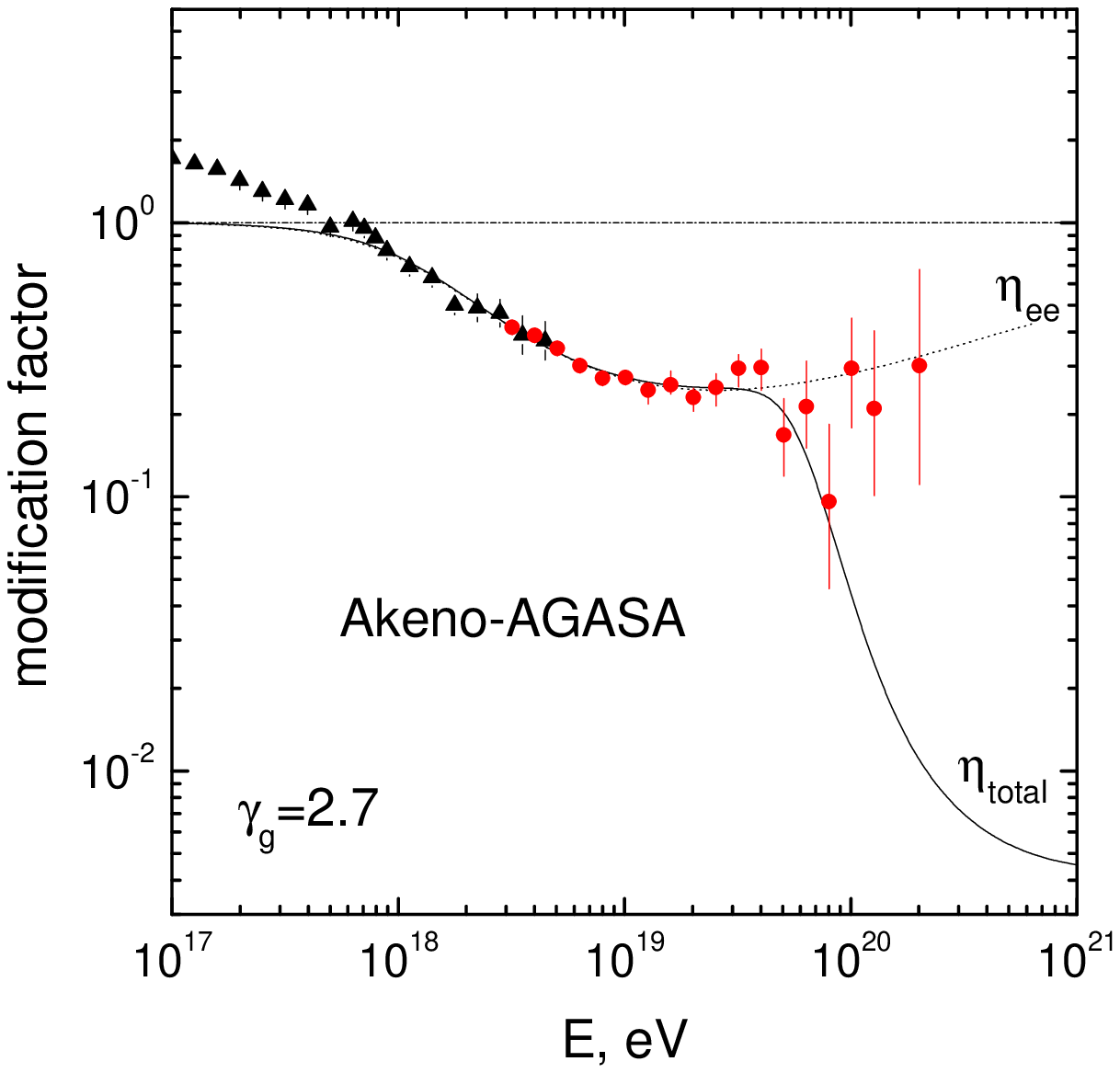}
   \end{minipage}
   \hspace{1mm}
   \vspace{-1mm}
 \begin{minipage}[h]{54mm}
    \centering
    \includegraphics[width=53mm]{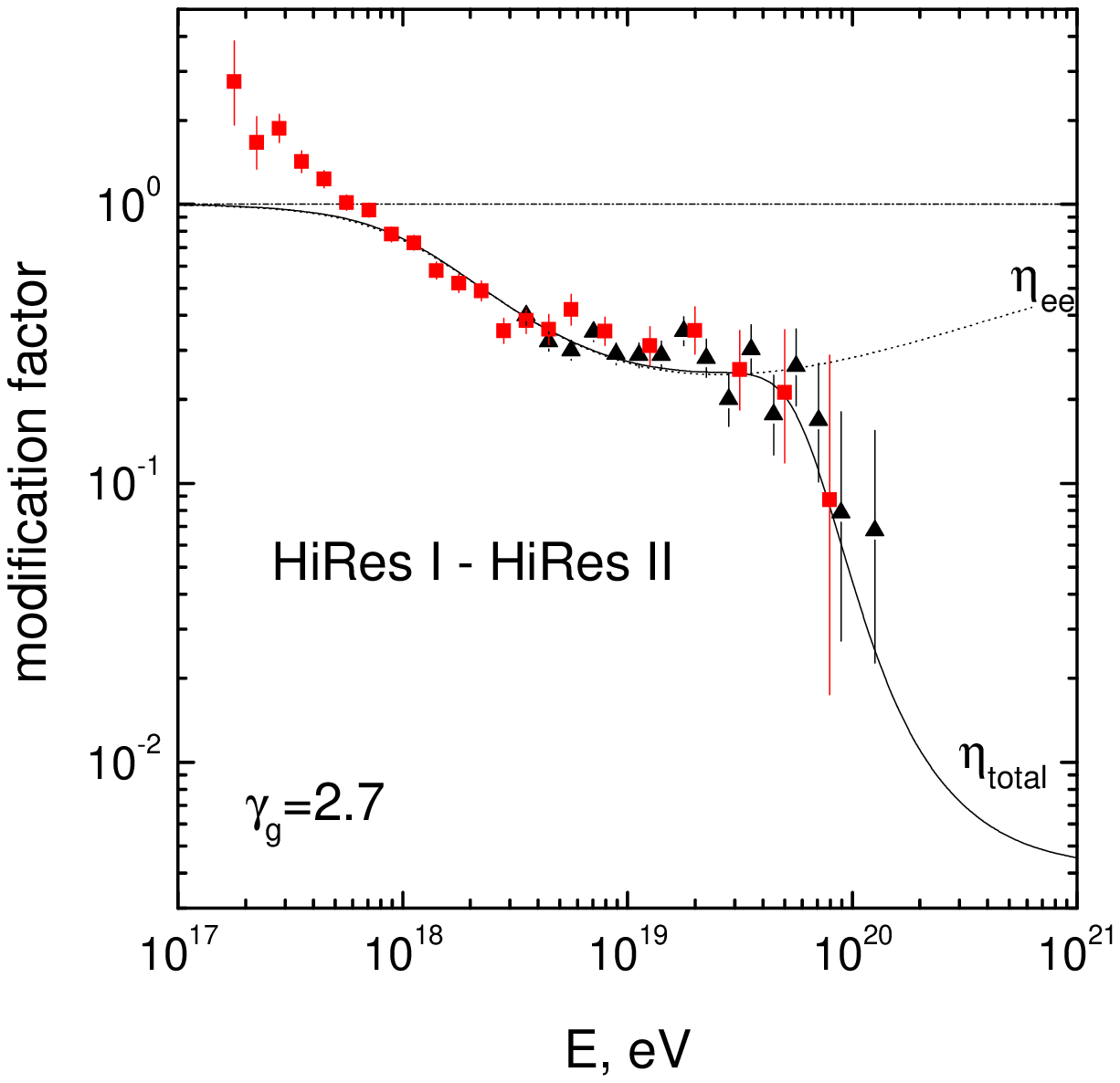}
 \end{minipage}
\medskip
   \begin{minipage}[ht]{54mm}
     \centering
     \includegraphics[width=53 mm]{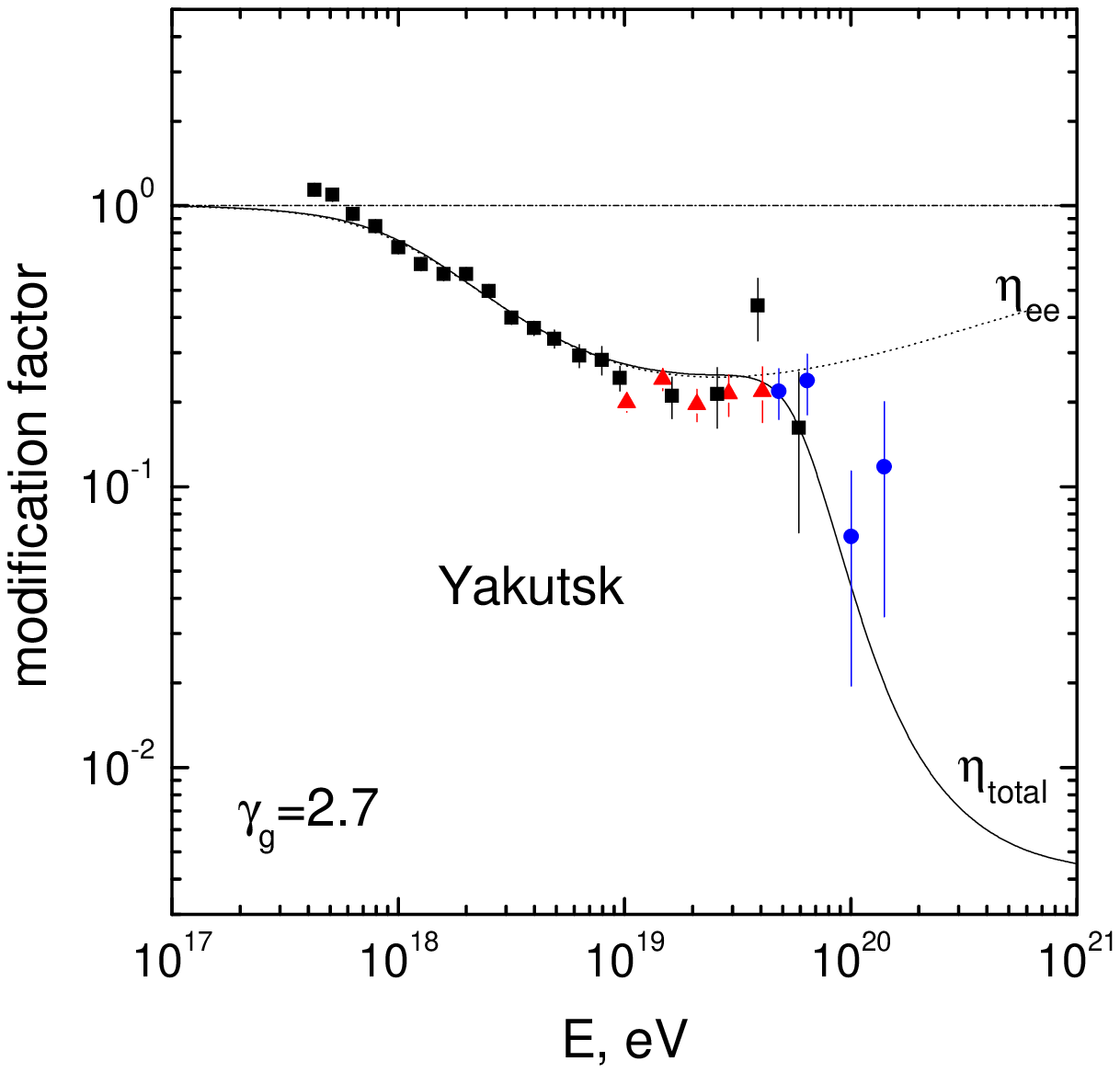}
   \end{minipage}
   \hspace{1mm}
   \vspace{-1mm}
 \begin{minipage}[h]{54mm}
    \centering
    \includegraphics[width=53mm]{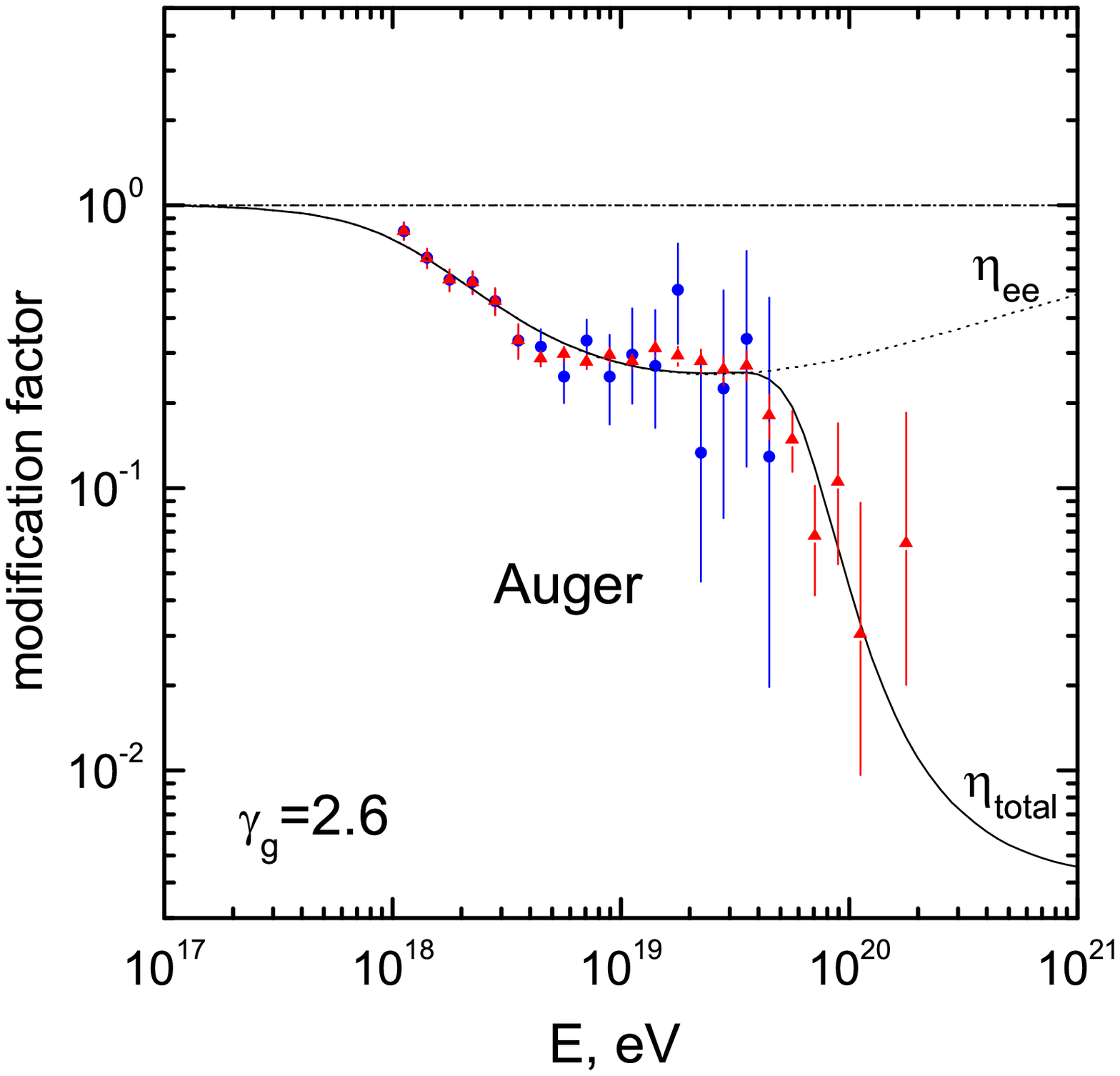}
 \end{minipage}
\end{center}
\vspace{-4 mm}%
\caption{ The predicted pair-production dip in comparison with
  Akeno-AGASA, HiRes, Yakutsk and Auger data.
  The first three experiments confirm dip with good
 $\chi^2/{\rm  d.o.f.} \approx 1.0 -1.2$, while the  Auger
 data are characterized
 by larger $\chi^2/{\rm d.o.f.}$ (see the text).
 The data of Fly's Eye  (not presented here) confirm
 the dip as well as  AGASA, HiRes and Yakutsk.} %
\label{fig:dips}
\end{figure}

The calculated modification factor is shown in comparison with
experimental data in Fig.~\ref{fig:dips}.
{\em As  Fig.~\ref{fig:dips} shows the pair production dip and
beginning of GZK cutoff up to energy $1\times 10^{20}$~eV is
reliably confirmed by all experimental data including AGASA}. As to
AGASA excess at $E > 1\times 10^{20}$~eV it can be explained by some
other  reasons, e.g.\ \cite{DBO2006} by systematic energy errors
combined with statistical fluctuations. The large $\chi^2$ in
comparison of the dip with Auger data is explained by energy errors
not included in the analysis.
\subsection{Evidence for GZK cutoff}
\begin{figure}[ht]
\centering
\includegraphics[height=6.0cm,width=8.0cm]{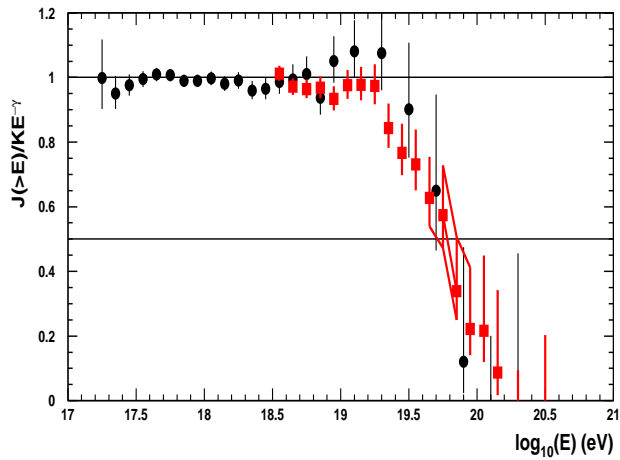}
\caption{$E_{1/2}$ as numerical characteristic of GZK cutoff in the
integral spectrum (see text).
 }
\label{fig:E-half}
\end{figure}
The Fig.~\ref{fig:dips} shows that data of HiRes, Auger and Yakutsk
are consistent with GZK cutoff. The presence of GZK cutoff is seen
most clearly in the HiRes data. However, low statistics and a
possibility of imitation of the observed steepening by other
reasons, e.g.\ by ``acceleration cutoff'', precludes one from making
the final conclusion. Recently, HiRes collaboration obtained the
numerical confirmation that this steepening is really the
GZK cutoff \cite{HiRes-GZK}. In the integral spectrum the GZK cutoff
is characterized by energy $E_{1/2}$, where calculated spectrum
$J(>E)$ becomes half of power-law extrapolation spectrum $K
E^{-\gamma}$ from low energies. As calculations \cite{BG} show this
energy is $E_{1/2} = 10^{19.72}$~eV for a wide range of generation
indices from 2.1 to 2.8. HiRes collaboration found $E_{1/2} =
10^{19.73 \pm 0.97}$~eV in a good agreement with the theoretical
prediction. In Fig.~\ref{fig:E-half} we reproduce the HiRes graph
from which $E_{1/2}$ was determined. The plotted value is given by
ratio of measured flux $J(>E)$ and its power-law approximation
$KE^{-\gamma}$. Extrapolation of this ratio to the higher energies
is given by 1, while intersection of measured ratio with horizontal
line 1/2 gives $E_{1/2}$.
\subsection{Calibration of detectors with help of dip and GZK
cutoff}
\begin{figure}[t]
\centering
\includegraphics[width=13.0cm]{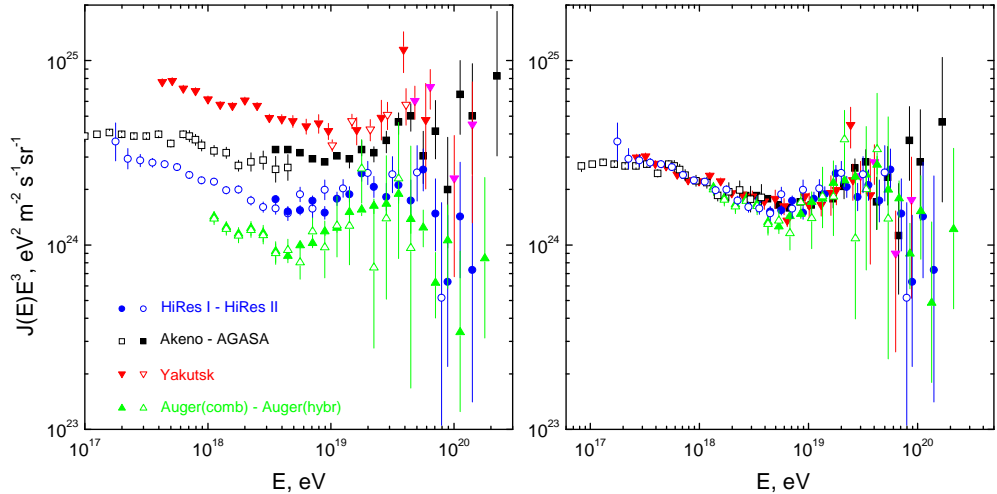}
\caption{The spectra and fluxes measured by Yakutsk, AGASA,
HiRes and Auger before (left panel) and after (right panel)
energy calibration.
}
\label{fig:calibration}
\end{figure}
The fluxes measured by Yakutsk, AGASA, HiRes and Auger are displayed
in Fig.~\ref{fig:calibration} (left panel).
To the large extent the great discrepancy seen there
is caused by comparison of the values
$E^3J(E)$ and thus systematic energy errors affect strongly this
contradiction. As was demonstrated in \cite{Be05,BGG-prd,Aloisio} the
energy calibration with help of the dip results in good agreement
between the absolute fluxes of all detectors. Here we use the
different approach for calibration of the detectors, based on both
features, dip and GZK cutoff. Since energies as measured by  HiRes
fit well the both features and especially GZK numerical
characteristic $E_{1/2}$, we assume that HiRes energies are correct
and the energies of all other detectors must be shifted by factor 
$\lambda$ to
reach the best agreement in fluxes (see also \cite{Kampert}). 
This procedure gives values of
$\lambda$ equal to 1.2, 0.75, 0.83, and 0.625 for Auger, AGASA,
Akeno and Yakutsk, respectively. This calibration does not give 
minimum $\chi^2$
for the dip shape, but describes better the dip and beginning of GZK
cutoff together. The fluxes after this
energy calibration are shown in Fig.~\ref{fig:calibration} (right
panel).
\subsection{Conclusions}
The dip and beginning of the GZK cutoff at $E \leq 1\times
10^{20}$~eV are signatures of proton interaction with CMB. They are
confirmed with good accuracy by all experimental data, including
AGASA. The GZK cutoff is seen most clearly in HiRes data and does
not contradict Auger and Yakutsk data. HiRes experiment has
numerical proof in the form of value $E_{1/2}$ that observed
steepening is really GZK cutoff. AGASA excess can be explained by
systematic energy errors combined with insufficient statistics. The
UHECR fluxes and spectra measured by all four detectors coincide
well after energy calibration of detectors with help of dip and GZK
cutoff. As recent Auger data \cite{Auger-AGN} indicate, the sources
of UHECR can be AGN (see \cite{BGG-AGN} for analysis).
\section{Cosmological Puzzles in the Past and Present}
The theoretical basis of cosmology is given by Friedmann solution of
the Einstein equation. After discovery of highly isotropic CMB
radiation and observational indications to the flat universe, it was
understood that Friedmann solution leads to two puzzles: {\em
horizon} and {\em flatness} problems.

The {\em horizon problem} can be explained in following way.

\begin{figure}[h]
\begin{minipage}[h]{7 cm}
\centering
\includegraphics[width=5.0cm]{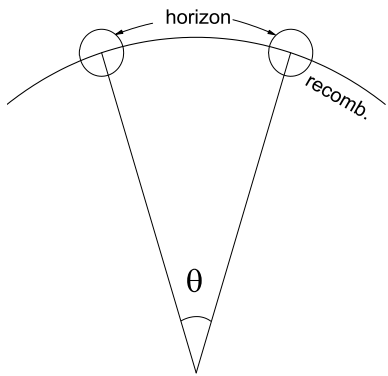}
\vspace{2mm} \caption{ Two regions on recombination sphere
($z=z_{\rm rec}$ separated by the horizon distance $ct_{\rm rec}$.)
} \label{fig:horizon}
\end{minipage}
\hspace{5mm}
\begin{minipage}[h]{8.2 cm}
CMB radiation decouples from matter after recombination time at
red-shift $z \approx 1100$ or $t_{\rm rec} \approx 1.2\times
10^{13}$~s. Two regions separated by horizon length $ct_{\rm rec}$
(see Fig.~\ref{fig:horizon}) are seen at angle $\theta \approx
(1+z_{\rm rec})ct_{\rm rec}/ct_0$, where $t_0 \approx 13.7$~Gyr is
the age of the universe. These regions are causally disconnected and
therefore the regions in the sky separated by angle $\theta \gsim
2^{\circ}$ cannot have equal CMB temperature. This conclusion is in
contradiction with observed isotropy of CMB radiation.

Why the universe is {\em flat} now?

Within Friedmann solution it can be only due to initial conditions
at $t_{\rm Pl} \sim 1/m_{\rm Pl}$. To have $\Omega - 1 \sim O(1)$
now, it is necessary to have at $t \sim t_{\rm Pl}$~~ $\Omega -1
\sim \xi$ with $\xi \sim 10^{-30}$.
\end{minipage}
\end{figure}

If $\Omega -1 >> |\xi|$ the universe collapses during time much
shorter than the age of universe now. If $\Omega -1 << -|\xi|$
galaxies and stars do not have enough time to be formed. Such
fine-tuning is highly unnatural.

The third cosmological puzzle was observation of rotational curves
in galaxies and cluster of galaxies, which showed that virial
(gravitational) masses of the objects are much bigger than visible
(luminous) mass. It was interpreted as dominance of weakly
interacting Dark Matter (DM). At present there are many particle
candidates for DM with detailed description how these particles can
be dominant in universe and produce Large Scale Structures (LSS).

As it is well known the first two puzzles were solved assuming an
early stage in expanding universe different from Friedmann regime:
{\em inflation}. At the different level of physical realization,
this stage was proposed in works by E. Gliner (1965) \cite{Gliner},
A. Starobinsky (1979) \cite{Star}, K. Sato (1981) \cite{Sato}, A.
Guth (1981) \cite{Guth}, A. Linde (1983) \cite{Linde}, A.Albrecht
and P. Steinhardt (1982) \cite{Stein}, and spectrum of density
perturbations (``galaxy formation'') imposed by this stage, was
first studied by S. Mukhanov and  G. Chibisov \cite{Mukhanov}.

Inflation is also expanding solution of the Einstein equation, but
driven by potential of the scalar field $\phi$,
inflaton. Expansion proceeds  exponentially, much faster than in
the Friedmann regime.

The Einstein equation for flat universe and energy conservation
result in the following equations \cite{W-book}:
\begin{eqnarray}
\dot{a}^2(t) &=& \frac{8\pi}{3} G a^2(t) \rho
\nonumber
\\
\ddot{a}(t) &=& -\frac{4\pi}{3} G (\rho +3 p) a(t)
\label{eq:Einstein}
\\
\dot{\rho}  &=& - 3H (p+\rho),
\nonumber
\end{eqnarray}
where $G$ is gravitation constant, $a(t)$ is the expansion factor of
the metric, $H(t)= \dot{a}/a$ is the Hubble parameter, $p$ and $\rho$
are pressure and density, connected by equation of state
$p=p(\rho)$. Let us consider the  matter with equation of state
$p=-\rho$ and $\rho(t) = \rho_0$, which follows from the third
equation above. Such equation of state in particular is realized by
scalar field $\phi$ rolling down in quasi-flat potential.
Eqs.~(\ref{eq:Einstein}) are easily solved in this case: %
\beq %
a(t) = a_0e^{H_0t} \;\; {\rm with}\;\; H_0^2 = \frac{8\pi}{3} G
\rho_0,
\label{eq:infl}
\eeq %
The solution (\ref{eq:infl}) describes
exponentially expanding bubble with constant Hubble parameter $H_0$.
This regime is called inflation.

Inflation obviously solves the horizon problem, because the size of
bubble $a(t)$ is much larger at any time $t$ than
the size of the observed universe and thus all observed regions are
causally connected.

The problem of flatness is also solved. To demonstrate it let us
generalize the first equation (\ref{eq:Einstein}) to the cases of
closed ($k=1$), open ($k=-1$) and flat ($k=0$) universes.
\beq %
\dot{a}^2(t) - \frac{8\pi}{3} G a^2(t) \rho= -k. \label{eq:Ein1}
\eeq %
Dividing this equation by $a^2$, and using $H=\dot{a}/a$ and
$\Omega=\rho/\rho_c$ we obtain \beq \Omega -1=\frac{k}{a^2H^2}.
\label{eq:flatness} \eeq For inflationary solution ($p=-\rho$),
$H=H_0$ and $a(t) \sim \exp(H_0 t)$ we see that at the end of
inflation, e.g.\ $t \sim  100 H_0^{-1}$ the r.h.s.\ is negligibly
small, providing the flatness at inflation and at all later universe
ages up to $t \sim t_0$.
\subsection{Where is Dark Matter?}
Observational cosmology is characterized now by very precisely
determined parameters, and the main contribution to accuracy of
their determination is given by WMAP data \cite{WMAP}. WMAP confirms
well $\Lambda$CDM  model with 6 parameters and the main parameters
have the following values: $H_0= 73.2$~km/sMpc,~ $\Omega_{\rm
tot}=1+\Omega_k$,~ $\Omega_k=-0.011 \pm 0.012$,~ $\Omega_b=
0.0416$,~ $\Omega_m=0.238$~, $\Omega_{\Lambda}=0.716$. However, the
precise determination of cosmological parameters are given by
combination of WMAP data with other observations such as SNI,
lensing, Large Scale Structures (LSS)  and
others. The most important results are determination of fraction of
Dark Matter (DM) and Dark Energy (DE). We summarize here the status of DM
according to these and other observations. 

{\em Direct search for DM} gives positive result only in observation
of modulation signal by DAMA \cite{DAMA}. It does not imply the direct
contradiction between DAMA and other experiments: the comparison is 
always model dependent and difference in target affects the comparison 
(see \cite{DAMA}). 

The strongest evidence for DM is
given by {\em indirect} methods. They include: 
\begin{itemize}
\item According to WMAP data density of matter $\Omega_m$ exceeds much
the baryon density $\Omega_b$ (see the data above). The presence of DM
is seen directly from WMAP data: without DM the height of the third 
acoustic peak would be much lower than observed.
\item Virial (gravitational) mass of galaxies and clusters of galaxies
is much larger than baryonic gas: $M_{\rm vir} \gg M_b$.
\item Theory of LSS formation (hierarchical clustering model)
  successfully explains the LSS formation using dominance of DM.
\end{itemize}

As was mentioned above only in DAMA the positive direct signal from DM 
(modulation) is observed.  Inspired by negative
results in other experiments, some authors
develop the alternative theories based on modified gravity. The
pioneering works in this direction, Modified 
Newtonian Dynamics (MOND), have been performed by Milgrom
\cite{MOND}. MOND is characterized by critical acceleration $a_0
\sim 1\times 10^{-8}$~cm s$^{-2}$. At $a <a_0$ the Newtonian
mechanics is modified, and the effects of DM are described by
modified gravity. The essential step forward in construction of a
theoretical model for modified General Relativity (GR) was made in
the work by Bekenstein \cite{Bekenstein}. He has build 
theoretically correctly the relativistic model of gravitation,
TeVeS, introducing three gravitational fields $g_{\mu\nu}$ (tensor,
like in GR), $U_{\mu}$ (vector) and $\phi$ (scalar). Apart from it
the model includes a  non-dynamical scalar $\sigma$, dimensional
constants $G$ and $l$, two dimensionless parameters $k$ and $K$ and
one arbitrary function $F(\sigma)$. The model tends to the 
standard GR in the proper limit and describes effects of MOND at low
acceleration. Thus, the model is much more complicated than the standard
GR and the only physical motivation of it is given by possibility to
avoid DM. This model successfully describes (with baryonic matter
only) flat rotation curves in galaxies, high velocities in clusters
and lensing. Recent work \cite{Dod} indicates that TeVeS may provide
the perturbation spectrum needed for LSS formation.

There are, however, two contradictions of MOND with observations:
\begin{enumerate}
  \item In the absence of DM the height of the third acoustic peak must
be much lower than observed by WMAP. In the paper \cite{WMAP} one
can find more general statement: ``Models without dark matter are
very poor fits to the data''.
  \item In the observed ``bullet cluster'' 1E0657-558 \cite{bullet}
(two colliding clusters) the gravitational potential is not centered
by X-ray emitting gas, which is the dominant baryon component in
this cluster (the ratio of the masses for gas and galactic
components is $M_{\rm gas}/M_{\rm gal} \sim 5 - 7$).
\end{enumerate}

In conclusion, I think that there is no DM puzzle. DM particles are
not seen in the directly-search experiments, either because
sensitivity is still low or because DM particles interact
superweakly with ordinary matter (e.g.\ gravitinos or SHDM
particles). MOND and TeVeS should be considered just as interesting
alternatives.
\subsection{Accelerated expansion of the universe}
\noindent As follows from the second equation (\ref{eq:Einstein})
for matter with equation of state $p = \omega \rho$ and with $\omega
< - 1/3$, acceleration $\ddot a > 0$, i.e.\ we have accelerating
expansion of metric. This is generalization of the inflation case
$\omega=-1$ which we considered above: $a(t) \propto \exp (H_0 t)$
and $\rho = \rho_0$ (vacuum energy). In the general case there can
exist both the matter described by energy-momentum tensor
$T_{\mu\nu}$ in the Einstein equation and vacuum energy described by
term  $\rho_{\rm vac}g_{\mu\nu}$. The latter is Lambda term, where
$\Lambda = 8\pi G \rho_{\rm vac}$.
In the general covariant form the Einstein equation reads
\beq %
R_{\mu\nu} - \frac{1}{2}g_{\mu\nu}R = -8\pi G (T_{\mu\nu}- \rho_{\rm
vac}g_{\mu\nu}).
\label{eq:Einst}
\eeq %
Presence of $\Lambda$ term
results in acceleration of expansion. While density of ordinary
matter in $T_{\mu\nu}$ diminishes with time, $\rho_{\rm vac}$
remains constant, and finally it dominates. Starting from this
moment universe enters again the phase of exponential expansion.

The presence of acceleration in the universe is established
experimentally in terms of non-zero $\Lambda$-term, but more
generally the data can be explained also by other terms which cause
accelerated expansion. For example, the SN data are explained by
non-standard dependence $H(z)$, where $H$ is the Hubble parameter.
Such dependence might appear due to matter with equation of state
$p=\omega \rho$ and $ -1 < \omega < -1/3$. This matter is usually
referred to as Dark Energy (DE) fluid.

In principle there are three sources which cause cosmic acceleration:
\begin{enumerate}
  \item [1.] Vacuum energy ($\Lambda$-term) in Eq.(\ref{eq:Einst}).
  \item [2.] DE fluid in $T_{\mu\nu}$ with equation of state $p=\omega \rho$
and $\omega < -1/3$. It can be realized as {\em quintessence},
ultra-light scalar field rolling down the potential; {\em phantom},
ghost field with equation of state given by $\omega < -1$; {\em
k-essence}; {\em Chaplygin gas} etc. (for review see
\cite{DE-rev,Wetterich}).
  \item [3.] Modified gravity. In this case the modified part of
Eq.~(\ref{eq:Einst}) is its l.h.s. No $\Lambda$-term or DE is needed
for acceleration and explanation of observational data. The most
interesting example is given by DGP \cite{Dvali} extra dimension
model, where gravity can leak into the bulk only at large distances.
This is one-parameter model which explains the observed
acceleration.
\end{enumerate}

Is it possible to distinguish between these possibilities by
observations?

Some options differ by spectrum of fluctuations and by equations of
state. WMAP data efficiently distinguish the models with different
$\omega$. The combined data of WMAP, LSS and SN gives $\omega =
-1.08 \pm 0.12$. With a prior assumption of flat universe the above
data give $\omega = -0.967 \pm 0.073$. It coincides well with
prediction of theory with $\Lambda$-term ($\omega = -1$) and is
allowed by some other models. Thus, the model with $\Lambda$-term
($\rho_{\rm vac}=$const) is allowed and preferable by its
simplicity.
\subsection{Vacuum energy problem}
Associated with $\Lambda$-term, the vacuum energy density is given
numerically as
\beq %
\rho_{\Lambda}=\Lambda/8\pi G= \Omega_{\Lambda}\rho_c= 4\times
10^{-47}~{\rm GeV}^4, \label{eq:vac-energy}
\eeq %
for $\Omega_{\Lambda}=0.73$ \cite{WMAP}. The vacuum energy
$\rho_{\rm \Lambda}$ can be presented as sum of energy density of
some cosmological field(s) $\sigma$  and  zero-modes of all known
fields (particles) $i$. Taking them as quantum oscillators with
ground-state energy $\omega_k/2$ one obtains vacuum energy of field
$i$
\beq %
\rho_{\rm vac}^{i} = \int_0^{k^i_{\rm max}} \frac{d^3k}{(2\pi)^3}
\frac{\omega_k}{2}. \label{eq:vac-i}
\eeq %
Then
the total vacuum energy is given by
\beq %
\rho_{\Lambda} =
\rho_{\sigma} + \sum_i \rho^i_{\rm vac}. \label{eq:vac-tot}
\eeq %
The problem is that the second term in r.h.s.\ of
Eq.~(\ref{eq:vac-tot}) is too large.  For example, the
reliably known and existing now in the universe  quark-gluon
condensate has vacuum energy between $-3\times 10^{-3}~{\rm GeV}^4$
as estimated in pioneering work \cite{SVZ} and $-0.01~{\rm GeV}^4$
in modern lattice QCD calculations. Its absolute value exceeds the
cosmological vacuum energy density (\ref{eq:vac-energy}) by 45
orders of magnitude (for further discussion see \cite{Dolgov1}).

The problem of too large zero-mode energy of ordinary fields in
comparison with cosmological constant is very old. It was already
discussed in late sixties (see \cite{Zeldovich}). It is important to
emphasize that this problem consists in cosmologically too large
vacuum energy of known fields and thus it is present for example in
all DE models. It may be a problem of elementary-particle physics.
For example with unbroken supersymmetry there is exact compensation
between negative vacuum energies of bosons and positive of fermions.
However, realistic breaking of this symmetry leaves the vacuum
energy too large. The 40 years of existing the problem 
were 40 years of intensive  brain attack ... and the problem
still exists. Among the proposed solutions there were the beautiful
ones. Among the authors there were the stars of theoretical physics
such as S. Coleman, S. Hawking, S. Weinberg, E.
Witten, Ya.\ Zeldovich. It makes me think that this puzzle will have
some unexpected solution.

S. Weinberg in his review ``The cosmological constant problem''
\cite{Weinberg} describes five roads which could lead to the
solution:
\begin{enumerate}
  \item[1.]  supersymmetry and supergravity,
  \item[2.]  anthropic approach,
  \item[3.] adjustment mechanisms,
  \item[4.] alternative gravity,
  \item[5.] quantum cosmology.
\end{enumerate}
All approaches are qualified as interesting and promising (quantum
gravity especially) but for all of them are indicated the weak
points, which do not allow to consider them as the correct, final
solution.

As the new direction we can add to the Weinberg list the
extra-dimension approach studied recently (for a review see
\cite{DE-rev}).
\subsection{Acceleration and Anthropic approach}
\begin{itemize}
  \item Why $\Lambda$-term is zero or very small?
  \item Why does acceleration start now?
  \item Why physical parameters are tuned to produce life?
\end{itemize}
These questions might have answers not in terms of physical
principles, but because in a universe with the ``wrong'' parameters
there is nobody to measure them. This is an anthropic solution. For
many interesting details I direct a reader to the exiting book of A.
Vilenkin ``Many Worlds in One'' \cite{V-book}.

The answer in the Anthropic approach to the above questions consists
in assumption that there is the infinite number of universes where
$\Lambda$ (or $\rho_{\rm vac}$) takes huge variety of different
values. In fact, these quantities do not have the fundamental
character to be the universal ones. The matter density $\rho_m$
diminishes with time faster than  $\rho_{\rm vac}$. Then if
$\Lambda$ is small (negative) a universe collapses early and
galaxies (observers) are not produced. If  $\Lambda$ is too large
the exponential expansion begins early and galaxies are not
produced, as well. The galaxy formation occurs roughly when
$\rho_{\rm vac} \sim \rho_m$. This is the basics of answering to the
first two questions and relevant calculations one can find in
\cite{pred}, where the value of $\Lambda$ was predicted.

The third question has a name ``coincidences''. To provide the life
many physical constants cannot be changed even by very small amount.
The famous example is given by F. Hoyle \cite{Hoyle}: the resonance
in reaction $3He^4 \rightarrow C^{12}$ is the most important channel
of carbon production in stars, without it we would not have life
(observer). There are many other coincidences, described e.g.\ in
\cite{coin}.
\subsection{From Inflation to Anthropic approach}
\noindent Anthropic approach very naturally appears in
chaotic-inflation scenario \cite{chaotic}. In this scenario the
distribution of the inflaton field $\phi$ in early universe is
assumed to be chaotic and respectively  the potential $V(\phi)$
is characterized by different values from very low ones up to
$M_{\rm Pl}^4$. In fact, this postulate of chaotic inflation is very
reasonable because different initial domains, separated by distance
larger than horizon, must have the different $\phi$ and $V$.  
Inflation is characterized by $H(\phi)=const$ for
each domain, where $\phi$ is quasi-homogeneous and it proceeds
independently in each horizon domain of size $H^{-1}(\phi)$, which
are called mini-universes. At present epoch $t=t_0$ the size of 
each mini-universe is many orders of magnitude larger than the  
horizon size $\sim 10^{28}$~cm.

{\em Since value of $\Lambda$ ($\rho_{\rm vac}$) is determined by
$\phi$ the different mini-universes are characterized by different
values of $\Lambda$. It provides a possibility of anthropic
selection of $\Lambda$.}

The chaotic inflation automatically results in the ``eternal
inflation'' \cite{eternal}, a process which has no beginning and
end. It occurs because the chaotic inflation includes two competing
processes of regeneration of inflationary regions and their decay.
There is some critical value $\phi_c$ below which $\phi$ oscillate,
potential energy is transferred to particles, and the Friedmann
expansion begins. On the other hand the region with relatively small
$\phi$ can through quantum tunneling turn into region of much larger
size. After tunneling the matter density $\rho_m$ becomes smaller, 
while $\rho_{\rm vac}$ remains the same. As a result 
$\rho_{\rm vac} > \rho_m$ provides the new round of inflation.  
In Ref.\cite{eternal} the process of increasing $\phi$ and 
$\rho_{\rm vac}$ is described as diffusion of $\phi$ to higher 
values as quantum fluctuation process. 
Thus the Universe always consists of exponentially large number
of mini-universities, part of which inflate and part experiences 
Friedmann expansion. They have the different values of $\Lambda$,
and therefore some of them might have an observer. 

All mini-universes have connected space-time. However,  there 
can be also the large or infinite number of universes with disconnected
space-time. This scenario is provided by ``creation of a universe 
from nothing''\cite{nothing}. 
A closed universe with zero total energy can be created by quantum
tunneling from initial object which has a vanishing size and mass.
This object can be considered as ``nothing''. If after-tunneling
object has the Planck size or less, it immediately collapses, if
more - it inflates, producing a disconnected  space-time 
and a universe. All these universes have different mini-universes 
with different $\Lambda$. 

\subsection{Status of anthropic approach}
Anthropic approach to the problem of vacuum energy is different 
from all other physics solutions and its status will depend on whether 
or not the irreproachable ordinary-physics solution (e.g. the 
compensation solution) will be found or not. I will remind again that
cosmological constant problem exists for more than 40 years,
attracting much attention of physics community. 
The negative attitude of large part of physics community to the 
anthropic approach could be conservatism of {\em present 
generation}, and some indication of it is given by discussion of 
anthropic solution  
by such deep thinkers of our generation as S. Weinberg, S. Hawking, 
M. Rees, A. Vilenkin, A. Linde and others. It could be that next 
generation of physicists will consider the anthropic approach as 
an ordinary physics. In fact, the anthropic approach reminds that 
of quantum mechanics. The quantum state of a universe   
with a fixed value of $\Lambda$ can be described by the wave 
function \cite{HH}, which gives the distribution of universes 
over different values of $\Lambda$.  In another version of this 
formalism \cite{Haw} each universe is characterised  not by single value
of $\Lambda$, but by their superposition, with the wave function
giving again a probability to find a universe with fixed value of $\Lambda$. 
The presence of observer in a universe (anthropic selection) plays 
the role similar to the the macro-detector in quantum mechanics.

\section*{Acknowledgements}
It is my great pleasure to thank Alex Vilenkin for helpful
discussions.


\bigskip

\begin{thebibliography}{99}
\bibitem{GZK}
Greisen K {\it Phys.\ Rev.\ Lett.\/} 1966 {\bf 16} 748; \\
Zatsepin G T and Kuzmin V A 1966 {\it Pisma Zh.\ Experim.\ Theor.\
Phys.\/} {\bf 4} 114

\bibitem{FE}
Bird D J \etal\ [Fly's Eye collaboration] 1994 {\it Ap.\ J.} {\bf
424}, 491


\bibitem{AGASA}
Shinozaki K \etal\ [AGASA collaboration] 2004 {\it Nucl.\ Phys.\/} B
(Proc.\ Suppl.) {\bf 136} 18

\bibitem{NaWa}
Nagano M and Watson A A 2000 {\it Rev.\ Mod.\ Phys.\/} {\bf 72} 689

\bibitem{BGZ}
Berezinsky V S, Grigorieva S I and Zatsepin G T 1975 {\it
Proc.\ 14th Int.\ Cosm.\ Ray Conf.\ (Munich)} {\bf 2} 711 \\
1976 {\it Izv.\ Acad.\ Nauk USSR (ser.\ phys.)} {\bf 40} 524

\bibitem{PSB}
Puget J L Stecker F W and Bredekamp J J 1976 {\it Astroph.\ J.} {\bf
205} 638

\bibitem{radio}
Berezinsky V S  1970 {\it Soviet Journ.\ Nucl.\ Phys.\/} {\bf 11} 399; \\
Prothroe R J and Biermann P 1966 {\it Astrop.\ Phys.\/} {\bf 6} 45

\bibitem{Be2000}
Berezinsky V 2000 {\it Nucl.\ Phys.\/} B (Proc. Suppl) {\bf 87} 387

\bibitem{Kirzh}
Kirzhnitz D A and Chechin V A 1972 {\it Pisma ZhETP} {\bf 14} 261; \\
1972 {\it Sov.\ Journ. of Nucl.\ Phys.\/} {\bf 15} 1051

\bibitem{BG}
Berezinsky V S and Grigorieva S I 1988 {\it Astron.\ Astroph.\/}
{\bf 199} 1

\bibitem{BGG-pl}
Berezinsky V, Gazizov A and Grigorieva S 2005 {\it Phys.\ Lett.\/} B
{\bf 612} 147

\bibitem{BGG-prd}
Berezinsky V, Gazizov A Z and Grigorieva S I 2006 {\it Phys.\
Rev.\/} D {\bf 74} 043005; hep-ph/0204357

\bibitem{Aloisio}
Aloisio R \etal\ 2007 {\it Astrop.\ Phys.\/} {\bf 27} 76

\bibitem{DBO2006}
De Marco D, Blasi P and Olinto A V 2006 {\it J. Cosm.\ Astrop.\
Phys.\/} {\bf 01} 002

\bibitem{HiRes-GZK}
Hires collaboration, arXiv:astro-ph/0703099

\bibitem{Be05}
Berezinsky V, astro-ph/0509069

\bibitem{Kampert}
Kampert K-H These Proceedings, astro-ph/0801.1986

\bibitem{Auger-AGN}
The Pierre Auger collaboration, astro-ph/0711.2256

\bibitem{BGG-AGN}
Berezinsky V, Gazizov A Z and Grigorieva S I, astro-ph/0210095

\bibitem{Gliner}
Gliner E B 1965 {\it Sov.\ Phys.\ JETP} {\bf 22} 378

\bibitem{Star}
Starobinsky A A 1979 {\it JETP Lett.\/} {\bf 30} 682

\bibitem{Sato}
Sato H 1981 {\it MNRAS} {\bf 195} 467

\bibitem{Guth}
Guth A H 1981 {\it Phys.\ Rev.\/} D {\bf 23} 347

\bibitem{Linde}
Linde A D 1982 {\it Phys.\ Lett.\/} B {\bf 108} 389

\bibitem{Stein}
Albrecht A and Steinhardt P J 1982 {\it Phys.\ Rev.\ Lett.\/} {\bf
48} 1220

\bibitem{Mukhanov}
Mukhanov V F and Chibisov G V 1981 {\it JETP Lett.\/} {\bf 33} 549

\bibitem{W-book}
Weinberg S 1972 {\it Gravitation and Cosmology, John Wiley and Sons}

\bibitem{WMAP}
Spergel D N \etal\ for WMAP collaboration 2007 {\it Ap.\ J. Suppl.}
{\bf 170} 377

\bibitem{DAMA}
Bernabei R \etal\ 2003 {\it  Riv.\ N.\ Cim.\/} {\bf 26} 1 ; 
2004 {\it Int.\ J.\ Mod.\ Phys.\/} D {\bf 13} 2127; 
2006 {\it Int.\ J.\ Mod.\ Phys.\/} A {\bf 21} 1445;
2007 {\it Int.\ J.\ Mod.\ Phys.\/} A {\bf 22} 3155;
2008 {\it Phys.\ Rev.\/} D {\bf 77} 023506

\bibitem{MOND}
Milgrom M 1983 {\it Ap.\ J.} {\bf 270} 365-84

\bibitem{Bekenstein}
Bekenstein J D 2004 {\it Phys.\ Rev.\/} D {\bf 70} 083509

\bibitem{Dod}
Dodelson S and Liguori M 2006 {\it Phys.\ Rev.\ Lett.\/} {\bf 97}
231301

\bibitem{bullet}
Clowe D \etal\ 2006 {\it Ap.\ J. Lett.\/} {\bf 648} L109

\bibitem{DE-rev}
Copeland E C, Sami M and Tsujikawa S 2006 {\it Int.\ J. Mod.\
Phys.\/} D {\bf 15} 1753

\bibitem{Wetterich}
Wetterich C, these Proceedings.

\bibitem{Dvali}
Dvali G G, Gabadadze G and Porrati M 2000 {\it Phys.\ Lett.\/} B
{\bf 495} 208

\bibitem{SVZ}
Shifman M, Vainstein A and Zakharov V 1979 {\it Nucl.\ Phys.\/} B
{\bf 147} 448

\bibitem{Dolgov1}
Dolgov A D 2005 {\it Proc.\ of LIX Yamada Conf.\ (eds. Suzuki H,
Yokoyama J, Suto Y and Sato K), Tokyo, Japan} 105

\bibitem{Zeldovich}
Zeldovich Ya.\ B 1967 {\it JETP Lett.\/} {\bf 6} 316

\bibitem{Weinberg}
Weinberg S 1989 {\it Rev.\ Mod.\ Phys.\/} {\bf 61} 1

\bibitem{V-book}
Vilenkin A 2006 {\it Many worlds in one, Hill and Wang, New York}

\bibitem{pred}
Weinberg S 1987 {\it Phys.\ Rev.\ Lett.\/} {\bf 59} 2607; \\
Vilenkin A 1995 {\it Phys.\ Rev.\ Lett.\/} {\bf 74} 846,
(gr-qc/9406010v2); \\
Carriga J and Vilenkin A, hep-th/0508005.

\bibitem{Hoyle}
Hoyle F 1953 {\it Ap.\ J.} {\bf 118} 513

\bibitem{coin}
Carr B J and Rees M J 1979 {\it Nature} {\bf 278} 605; \\
Davies P C W 1982 {\it The accidental universe (Cambridge University
Press, Cambridge)}

\bibitem{chaotic}
Linde A D 1983 {\it Phys.\ Lett.\/} B {\bf 129} 177

\bibitem{eternal}
Vilenkin A 1983 {\it Phys.\ Rev.\/} D {\bf 27} 2848; \\
Linde A D 1986 {\it Phys.\ Lett.\/} B {\bf 175} 395

\bibitem{nothing}
Zeldovich Ya\ B  1981 {\it Sov.\ Astron.\ Lett.\/} {\bf 7} 322; \\
Vilenkin A 1982 {\it Phys.\ Lett.\/} {\bf 117} 25

\bibitem{HH}
Hartle J B and Hawking S W 1983 {\it Phys.\ Rev.\/} D {\bf 29} 2960

\bibitem{Haw}
Hawking S W 1987 {\it Phys.\ Scr.\/} T {\bf 25} 202


\end{thebibliography}
\end{document}